# Role of Magnon-Magnon Scattering in Magnon Polaron Spin Seebeck Effect

Zhong Shi,[1, 2,*] Qing Xi,[1] Junxue Li,[2] Yufei Li,[1] Mohammed Aldosary,[2] Yadong Xu,[2] Jun Zhou,[3, **] Shi-Ming Zhou,[1] and Jing Shi[2, †]

[1]Shanghai Key Laboratory of Special Artificial Microstructure and School of Physics Science and Engineering, Tongji University, Shanghai 200092, China

[2]Department of Physics and Astronomy, University of California, Riverside, California 92521, USA

[3]NNU-SULI Thermal Energy Research Center & Center for Quantum Transport and Thermal Energy Science, School of Physics and Technology, Nanjing Normal University, Nanjing 210023, China

(Revision date: October 11, 2021)

The spin Seebeck effect (SSE) signal of magnon polarons in bulk-$Y_3Fe_5O_{12}$ (YIG)/Pt heterostructures is found to drastically change as a function of temperature. It appears as a dip in the total SSE signal at low temperatures, but as the temperature increases, the dip gradually decreases before turning to a peak. We attribute the observed dip-to-peak transition to the rapid rise of the four-magnon scattering rate. Our analysis provides important insights into the microscopic origin of the hybridized excitations and the overall temperature dependence of the SSE anomalies.

Corresponding authors:

Zhong Shi: shizhong@tongji.edu.cn, Jun Zhou: zhoujunzhou@njnu.edu.cn and Jing Shi: jing.shi@ucr.edu

Z.S. and Q.X. contributed equally to this work.

Spin Seebeck effect (SSE), an important spin caloritronic phenomenon discovered in 2008, has attracted much attention in the spintronics and condensed matter physics communities [1–7]. In the SSE, which is regarded as a spin analog of the Seebeck effect, spin accumulation is induced in a ferromagnetic (FM) layer by a temperature gradient. Through $s$-$d$ scattering at the FM/non-magnetic metal (NM) interface, spin current is generated in the NM and subsequently detected via the inverse spin Hall effect. Apparently, the SSE in the FM/NM heterostructures enables a simple and versatile way of generating spin current from heat, and thus is of great importance for spintronic applications. Furthermore, the SSE provides a new way of enhancing the thermoelectric efficiency [5].

The SSE strongly depends on the propagation of magnons, quanta of spin waves, in the FM layer, which is determined by the excitation and the scattering of magnons [5]. Especially, the SSE can be influenced by strong magnon-phonon coupling, as observed in the pioneering work by Kikkawa *et al* [8]. The dispersion of magnons in FM can be modified by the external magnetic field $H$ as follows

$$\omega_k = \sqrt{\gamma\mu_0 H + D_{ex}k^2} \\ \times \sqrt{\gamma\mu_0 H + D_{ex}k^2 + \gamma\mu_0 M_S \sin^2\theta_k}, \tag{1}$$

where $M_S$ is the saturation magnetization, $\gamma$ the gyromagnetic ratio $\gamma = g\mu_B/\hbar$, $g$ the spectroscopic factor, $\mu_B$ the Bohr magneton, $\hbar$ the reduced Planck constant, and $D_{ex}$ the exchange stiffness [9]. The applied magnetic field $H$ is set along the $z$-axis and $\theta_k$ is the angle between the wave vector $\boldsymbol{k}$ and $H$ [see inset of Fig. 1(a)]. On the other hand, with the long wavelength approximation, the dispersion of transverse(longitudinal) acoustic phonons can be expressed as $\omega_q = qc_{T(L)}$, where $c_T$ and $c_L$ are the sound velocities of the transverse acoustic (TA) and longitudinal acoustic (LA) phonons, respectively.

At low external magnetic fields, magnon and TA phonon dispersion curves intersect with each other twice, as shown by the solid black and green curves in Fig. 1(b). With increasing $H$, the dispersion curve of magnons is lifted vertically, represented by the black dashed line. When the magnon dispersion becomes tangent to TA phonon dispersion at the touching magnetic field $H_{TA}$, the two crossing points merge into one. The same situation happens to the magnon and LA phonon dispersion curves at $H = H_{\text{LA}}$. If the magnon-phonon interaction is too strong to be considered as a perturbation, magnon polarons are formed through magnon-phonon hybridization at the tangent points, where their momentum $(k, q)$ and energy $(\varepsilon)$ are matched [see Figs. 1(a) and 1(b)].

When the magnon polarons are formed, the SSE voltage could be enhanced or suppressed. Consequently, resonant-like SSE anomalies appear near the tangent points at $H_{\text{TA(LA)}}$. Theoretical studies have shown that the magnon polaron SSE can be enhanced or suppressed with respect to the ordinary magnon SSE background, depending on whether the scattering rate of magnons ($\tau_{\text{mag}}^{-1}$) is larger or smaller than that of phonons ($\tau_{\text{ph}}^{-1}$) [9]. To date, most studies have focused on the scattering processes of magnons and phonons by impurities. Accordingly, at the tangent point, the ratio of magnon- and phonon-impurity scattering rates $\eta = \tau_{\text{mag,impurity}}^{-1} / \tau_{\text{ph,impurity}}^{-1}$ does not change with temperature because both the magnon and phonon impurity scattering rates are independent of temperature. If the impurity scattering dominates [8,10,11], either peaks or dips are superimposed on the SSE background signal over the entire temperature range, for $\eta > 1$ or $\eta < 1$, respectively. However, at finite temperatures, three-magnon and four-magnon scattering processes are activated so that $\eta$ can depend on temperature. Indeed, the magnon-magnon scattering can play important roles in the propagation of magnons [12,13], magnon Bose-Einstein condensation [14], and Raman scattering [15]. In this Letter, we report that in bulk-$Y_3Fe_5O_{12}$ (YIG) with moderately high acoustic quality, the magnon polaron SSE evolves from dip to peak as temperature increases, which provides direct evidence of magnon-magnon scattering especially the four-magnon scattering contribution to the SSE.

The heterostructures used in this study consist of 0.5 mm-thick polished single crystal YIG substrates and 5 nm-thick Pt layers. The YIG crystals are provided by MTI Corporation and the University of Texas at Austin, referred as MTI-YIG and UTA-YIG, respectively. The SSE results included in the main paper are from the MTI-YIG sample and those from the UTA-YIG can be found in the Supplementary Material (SM), Sec. IV [16]. Pt is used as a spin current detector. The details of sample fabrication and transport measurement are described elsewhere [16].

A typical SSE loop in MTI-YIG/Pt is shown in Fig. 1 (c). This 50 K loop shows a nearly linear background at high magnetic fields, which is generally attributed to the suppressed magnon population [22,29]. The SSE voltage is defined as $V_{\text{SSE}} = \frac{1}{2}[V(+H) - V(-H)]$. $V_{\text{SSE}}$ is about 28.5 $\mu$V at 50 K for an ac heating current of 2.5 mA$_{rms}$ at $\mu_0 H$ =2.5 T. Notably, four kink-like anomalies can be identified around $\mu_0 H = \pm 2.5$ T and $\pm 9.5$ T, which are enlarged in Figs. 1(d) and 1(e). The magnon polaron-induced SSE anomaly is characterized by $\Delta V_{\text{TA(LA)}-\text{SSE}} = V_{\text{dip/peak}} - V_{\text{base}}$ for the TA(LA) magnon polaron. Clearly, at 50 K, both $\Delta V_{\text{TA}-\text{SSE}}$ and $\Delta V_{\text{LA}-\text{SSE}}$ are negative at $H_{\text{TA}}$ and $H_{\text{LA}}$, respectively, with the magnitude of $\Delta V_{\text{LA}-\text{SSE}}$ being slightly larger than that of $\Delta V_{\text{TA}-\text{SSE}}$. In the following, we will focus on the TA magnon polaron SSE, which exhibits a stronger temperature dependent behavior. The discussion about the LA magnon polaron SSE can be found in SM, Sec. III [16].

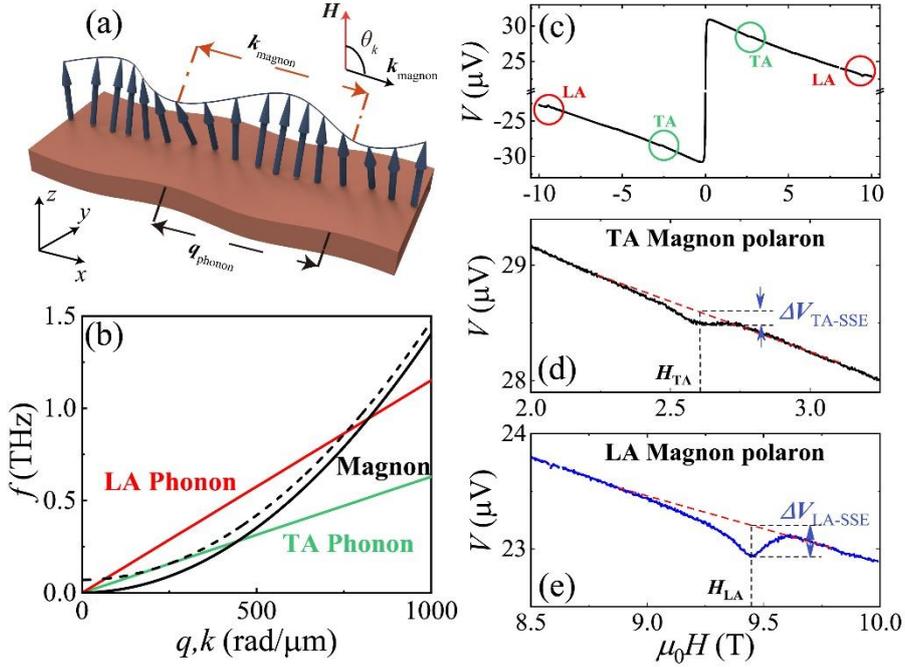

**Figure 1:** (a) Sketch of spin wave ($k_{\text{magnon}}$) and acoustic wave ($q_{\text{phonon}}$) as they are simultaneously excited in YIG with the same group velocity. The inset shows the sketch of external field $H$, wavevector $\boldsymbol{k}$ and $\theta_k$. (b) Dispersions of TA (green) and LA (red) phonons, and magnons for $H = 0$ (solid black) and $H = H_{TA}$ (dashed black). (c) SSE loop of MTI-YIG/Pt with $I_{\text{heater}}$=2.5 mA$_{rms}$ at $T$=50 K. The anomalies of TA and LA magnon polarons are marked by colored circles. (d) and (e) show enlarged regions near $H_{\text{TA}}$ and $H_{\text{LA}}$. The TA(LA) magnon polaron SSE anomaly is defined as $\Delta V_{\text{TA(LA)-SSE}} = V_{\text{dip}} - V_{\text{base}}$ and $H_{\text{TA}}$ ($H_{\text{LA}}$) is indicated.

In order to investigate contributions from other mechanisms than impurity scattering, we have measured the TA magnon polaron SSE in the YIG/Pt sample from 10 to 300 K. Interestingly, the shape of the SSE anomaly undergoes a clear transition, as shown in Fig. 2(a). At low temperatures, a dip is observed and its magnitude decreases with increasing temperature. It disappears at ~250 K, but a peak begins to appear at higher temperatures. Namely, $\Delta V_{\text{TA}-\text{SSE}}$ exhibits a sign change over the temperature range from 10 K to 300 K, distinctly different from the previous results reported by Kikkawa and Ramos [8,11], in which the same sign was observed up to 300 K. As indicated by the red dashed line in Fig. 2(a), $H_{\text{TA}}$ increases gradually as temperature decreases, which is summarized in Fig. 2(b).

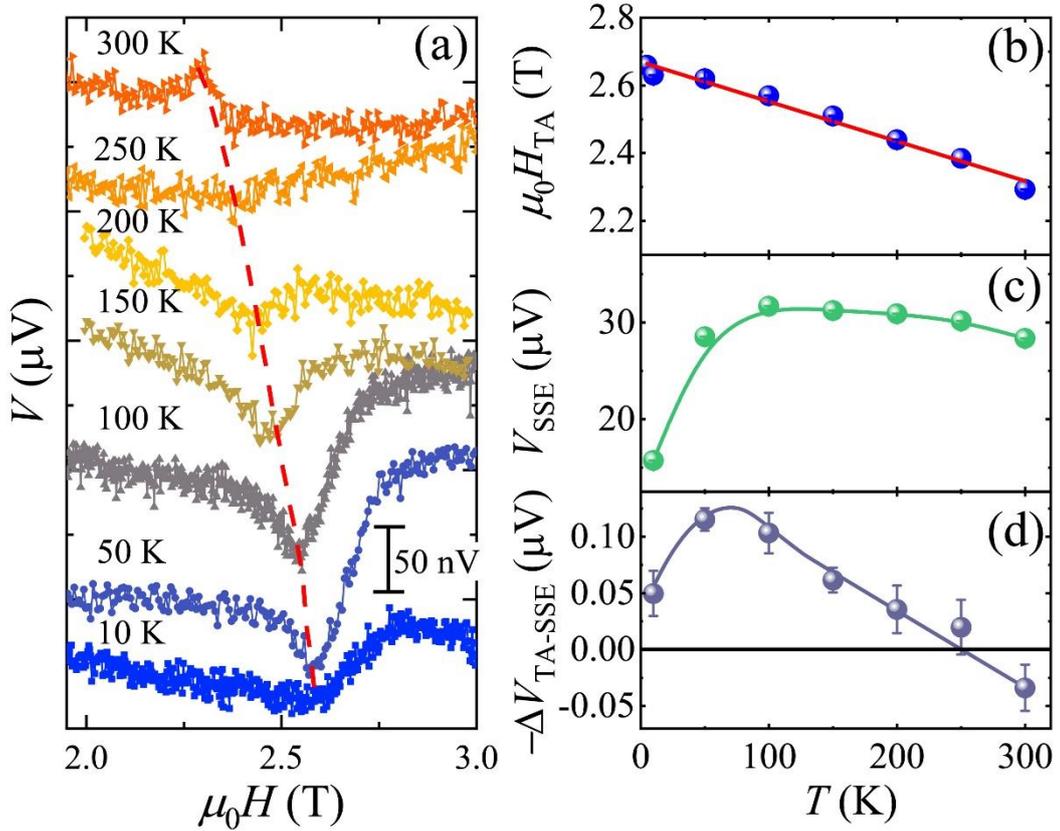

**Figure 2:** (Left) (a) SSE voltages near $H_{\text{TA}}$ at various temperatures. (Right) $\mu_0 H_{\text{TA}}$ (b), SSE voltage $V_{\text{SSE}}$ (c), and SSE change $\Delta V_{\text{TA-SSE}}$ (d) vs. temperature. Linear SSE background in (a) is subtracted and all curves are vertically shifted for clarity. The red solid line in (b) is a linear fit. The dashed line in (a), solid lines in (c) and (d) are guides to the eye.

The overall temperature dependent behavior of the TA magnon-polaron anomaly is presented in the right panel of Fig. 2. Fig. 2(b) shows the temperature dependence of $\mu_0 H_{\text{TA}}$. It decreases linearly from 2.67 T at 10 K to 2.29 T at 300 K, which is in contrast with the nonmonotonic behaviors of $V_{\text{SSE}}$ and $\Delta V_{\text{TA}-\text{SSE}}$ over the same temperature range, as shown in Figs. 2(c) and 2(d). The $H_{TA(LA)}$ can be expressed in Eq. 2 below under the approximation of $H >> M_S$ [9],

$$\mu_0 H_{TA(LA)} = \frac{c_{T(L)}^2}{4\gamma D_{ex}}. \qquad (2)$$

$D_{ex}$ of YIG is found to be nearly constant below 300 K as discussed in SM, Sec. I [16]. According to Eq. 2, this small change cannot account for the observed 15% decrease in $H_{TA}$ from 10 K to 300 K. However, the sound velocity in YIG decreases by 7% over the same temperature range reported by Cornelissen *et al*, which can produce a 14% decrease in $H_{TA}$ [10]. Moreover, since the Young's modulus is three times as large as the rigidity modulus in YIG, it leads to $c_L^2 \sim 3c_T^2$ [30], which can result in a three-fold difference between $H_{LA}$ and $H_{TA}$ according to Eq. 2. This is consistent with our observation that $H_{LA}$ is approximately three times as large as $H_{TA}$, as shown in Figs. 1(c), 1(d), and 1(e). Therefore, the overall reduction in $H_{TA}$ is mainly due to the phonon velocity change.

Now let us compare the temperature dependences of $V_{SSE}$ and $\Delta V_{TA-SSE}$ as shown in Figs. 2(c) and 2(d). The non-monotonic characteristics in $V_{SSE}$ and $-\Delta V_{TA-SSE}$ with maxima near 100 K are similar to the results of Kikkawa *et al* [29]. In general, $V_{SSE}$ in ferromagnets is caused by either magnon or phonon heat flow [31–34], but they have quite different consequences. The former mechanism produces a monotonic temperature dependence [32], whereas the latter gives rise to a pronounced peak at low temperatures. Therefore, the non-monotonic temperature dependence of $V_{SSE}$ in Fig. 2(c) can be mainly attributed to phonon drag [33–36]. At low temperatures, the phonon density is small based on the Bose-Einstein distribution function, whereas the relaxation time $\tau_{ph}$ is large. As the temperature increases, $\tau_{ph}$ deceases more steeply than phonon density increases because of the Umklapp scattering. These opposing trends result in a maximum in spin current density and thus the SSE peak at some intermediate temperature. The detailed temperature dependence of phonon and magnon relaxation times for different scattering processes will be discussed later.

Fig. 2(d) is the temperature dependence of the magnon polaron anomaly in reference to the smooth SSE background, with the positive (negative) sign denoting a peak (dip). This anomaly arises from the strong interaction between phonon and magnon at the tangent point. Since the magnon-phonon interaction is too strong to be treated as a perturbation, pure magnon or phonon dispersion no longer exists. Instead, it is more appropriate to treat them as magnon polarons considering the effect of spin-lattice coupling on magnon and phonon dispersion. Therefore, the spin Seebeck coefficient ($\zeta_{xx}$) can be derived from the general transport theory as shown in Eq. 3, if only impurity scattering of phonons and magnons is considered, and the contribution from $k = k_{TA}$ modes is approximately expressed as a function of $\delta H$ around $H_{TA}$ ($\delta H \to 0$) [9],

$$\zeta_{xx} \sim \frac{\beta}{TL^2|V_{\text{mag}}|^2}(\partial_k\Omega_{1k})^3\frac{e^{\beta\hbar\Omega_{1k}}}{(e^{\beta\hbar\Omega_{1k}}-1)^2}(\hbar\Omega_{1k}) \\ \times y_1^{mag}(\delta H,\eta), \tag{3}$$

where $\Omega_{i\mathbf{k}}$ is the frequency of the $i$th magnon-polaron mode, $\beta = 1/k_B T$, $L$ the thickness of YIG film, $V_{\text{mag}}$ the magnon-impurity scattering potential, and $y_1^{mag}$ is

$$y_1^{\text{mag}}(\delta H,\eta) = \frac{2\eta\left[1 + 2(\tilde{k}\delta H)^2 + \eta\right]}{1 + \eta\left[2 + 4(\tilde{k}\delta H)^2 + \eta\right]}, \tag{4}$$

with $\tilde{k} = \mu_0\gamma/\sqrt{4Sk}$, and $\eta = |V_{\text{mag}}/V_{\text{ph}}|^2 = \tau_{\text{mag,impurity}}^{-1}/\tau_{\text{ph,impurity}}^{-1}$. When $\delta H = 0$, $y_1^{\text{mag}}(0,\eta) = \frac{2\eta}{1+\eta}$, and the factor in front of $y_1^{mag}$ corresponds to the extrapolated $V_{\text{SSE}}$ background at $H_{\text{TA}}$. From Eq. 3, we obtain that the relative SSE change at the tangent point $\Delta\zeta_{xx}/\zeta_{xx}$ is approximately proportional to $y_1^{\text{mag}}(\eta) - y_1^{\text{mag}}(\eta)|_{\eta=1} \sim (\eta - 1)/2$ for $\eta$ close to 1. Thus, the magnon polaron SSE is expected to be enhanced for $\eta > 1$ and suppressed for $\eta < 1$ (see SM, Sec. VI [16]).

In order to further explain the results in Figs. 2(c) and 2(d), it is necessary to include other scattering mechanisms and reevaluate $\eta$ accordingly, especially to calculate the total scattering rates of magnon and phonon *as a function of temperature*. The ratio $\eta$ between the total scattering rates of magnon and phonon could be expressed as $\eta = \tau_{\text{mag}}^{-1}/\tau_{\text{ph}}^{-1}$ without loss of generality in the following discussion. We calculate the total scattering rate of phonons as,

$$\tau_{ph}^{-1} = \tau_{ph,boundary}^{-1} + \tau_{ph,impurity}^{-1} + \tau_{ph,Umklapp}^{-1}, \tag{5}$$

where $\tau_{ph,boundary}^{-1}$ and $\tau_{ph,impurity}^{-1}$ are the scattering rate of phonon-boundary and impurity scattering, respectively. $\tau_{ph,Umklapp}^{-1}$ is the scattering rate of the Umklapp phonon-phonon process. Both the phonon-boundary and impurity defects contribute to the temperature-independent scattering rate, while the Umklapp process leads to a $T$-dependent scattering rate. When three-magnon and four-magnon scattering processes are considered, the total scattering rate of magnons is expressed as follows,

$$\tau_{mag}^{-1} = \tau_{mag,boundary}^{-1} + \tau_{mag,impurity}^{-1} + \tau_{mag,3\text{-magnon}}^{-1} + \tau_{mag,4\text{-magnon}}^{-1}, \tag{6}$$

where $\tau_{mag,boundary}^{-1}$, $\tau_{mag,impurity}^{-1}$, $\tau_{mag,3\text{-magnon}}^{-1}$, and $\tau_{mag,4\text{-magnon}}^{-1}$ are scattering rates of boundary scattering, magnon-impurity scattering, three-magnon scattering, and four-magnon scattering, respectively. Apparently, the defect scattering rates, $\tau_{mag,boundary}^{-1}$ and $\tau_{\text{mag,impurity}}^{-1}$, are independent of temperature. The three-magnon and four-magnon scattering rates due to the dipole-dipole and exchange interactions scale as $T$ and $T^{5/2}$, respectively. The detailed theoretical calculations are presented in SM, Sec. VI [16].

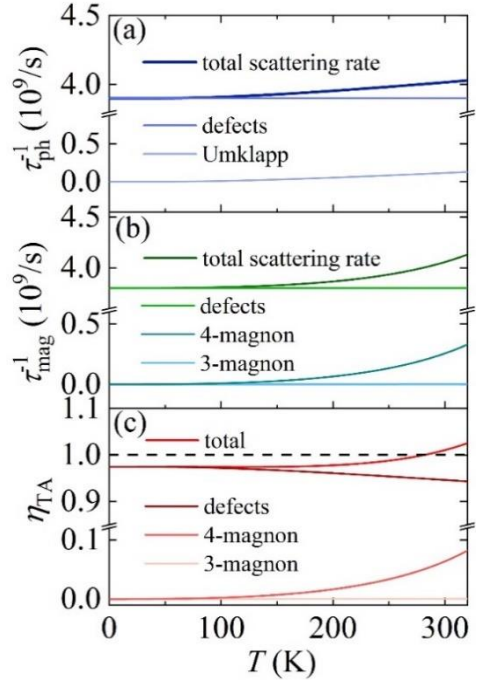

**Figure 3:** Calculated TA phonons (a) and magnons (b) scattering rates as functions of temperature. The four curves in (c) represent the ratios $\eta_{TA}$ of the total magnon scattering rate $\tau_{mag}^{-1}$ and three individual magnon scattering rates to the total TA phonon scattering rate $\tau_{ph}^{-1}$. The elastic scattering component by defects includes contributions from the temperature-independent boundary- and impurity-scattering processes. All parameters were taken from Table SII in SM [16].

Calculated $\tau_{\text{ph}}^{-1}$, $\tau_{\text{mag}}^{-1}$ and $\eta_{\text{TA}}$ at the tangent point [ $k = q = c_{\text{T}}/(2D_{\text{ex}})$ and $H = H_{\text{TA}}$ ] as functions of temperature are plotted in Fig. 3. At high temperatures, more high energy phonons are activated and the scattering rate of the Umklapp phonon-phonon process becomes significant, which scales linearly with $T$. Meanwhile, the scattering rates $\tau_{\text{mag,3-magnon}}^{-1}$ and $\tau_{\text{mag,4-magnon}}^{-1}$ due to three-magnon and four-magnon processes follow the $T$ and $T^{5/2}$ dependences, respectively. Therefore, the latter rises more steeply than the former at high temperatures. The four curves in Fig. 3(c) represent the ratios of the total magnon scattering rate $\tau_{\text{mag}}^{-1}$ and three individual magnon scattering rates to the total phonon scattering rate. Clearly, the increasing four-magnon scattering rate surpasses the phonon scattering rate at high temperatures. Consequently, it is the four-magnon scattering process that drives $\eta >$ 1.0 at high temperatures, producing a sign change of the magnon polaron SSE. This is in good qualitative agreement with the experimental observation [Fig. 2(d)]. Moreover, the calculated temperature corresponding to $\eta = 1$, i.e., the SSE sign change temperature, for TA mode is slightly lower than that for LA mode as shown in Fig. S13, resulting in a lower sign change temperature of $\Delta V_{\text{TA-SSE}}$ than that of $\Delta V_{\text{LA-SSE}}$. As shown in Fig. S5, no LA anomaly sign change is observed up to 250 K where the TA sign change occurs, which is consistent with the calculations.

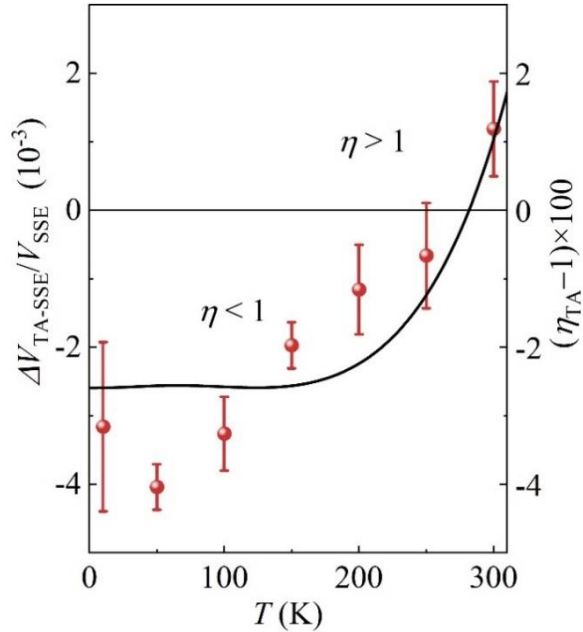

**Figure 4:** Temperature dependencies of calculated $\eta_{\text{TA}} - 1$ (solid line) and measured $\Delta V_{\text{TA-SSE}}/V_{\text{SSE}}$ (circles).

In order to make a more quantitative comparison between the observed and calculated temperature dependences of $\Delta V_{\text{TA-SSE}}$, we plot $\Delta V_{\text{TA-SSE}}/V_{\text{SSE}}$ over the entire temperature range. Since both $\Delta V_{\text{TA-SSE}}$ and $V_{\text{SSE}}$ depend on the thermal conductivity of YIG and spin-charge conversion in the same way, the ratio $\Delta V_{\text{TA-SSE}}/V_{\text{SSE}}$ does not contain any of these quantities and their temperature dependences. Therefore, $\Delta V_{\text{TA-SSE}}/V_{\text{SSE}}$ no longer depends on the thermal conductivity (see SM, Sec. V), . According to Eqs. 3 and 4, we have $\Delta V_{\text{TA-SSE}}/V_{\text{SSE}} \propto (\eta_{\text{TA}} - 1)/2$ at the tangent point. Fig. 4 shows the comparison between $\Delta V_{\text{TA-SSE}}/V_{\text{SSE}}$ and $(\eta_{\text{TA}} - 1)$ for the TA SSE anomaly. Despite the experimental uncertainty, they show very good agreement with each other, especially on the sign change. It is important to note that this model explains why $\Delta V_{\text{TA-SSE}}$ changes more sharply than $V_{\text{SSE}}$ at high temperatures, as shown in Figs. 2(c) and 2(d), due to the extra $y_1^{\text{mag}}$-factor for the magnon polaron SSE.

Based on our model, the magnon polaron SSE behavior can be classified into the following three main regimes according to low-temperature parameter $\eta$ value. (1) $\eta > 1.0$, i.e., high acoustic quality or weak phonon defect scattering. The magnon polaron SSE increases with increasing temperature because the four-magnon scattering rate increases more rapidly than that of the Umklapp phonon-phonon process. Hence, $\eta$ is always larger than 1.0 at all temperatures. Accordingly, the magnon polaron SSE peak always prevails and its magnitude increases with increasing temperature. This is the case in the experiment reported by Kikkawa [8] and our own experiment as shown in Fig. S7 [16]. (2) $\eta \ll 1.0$, i.e., low acoustic quality. $\eta$ is always smaller than 1.0 below *300* K. Accordingly, the magnon polaron SSE dip feature always prevails and its magnitude decreases with increasing temperature. The nonlocal SSE in heterostructures consisting of YIG grown by liquid phase epitaxy [10] belongs to this regime. (3) $\eta$ smaller than but close to 1.0, i.e., moderately high acoustic quality. Owing to the rapidly rising four-magnon scattering rate, $\eta$ becomes larger than 1.0 at high temperatures; therefore, the $\Delta V_{\text{TA-SSE}}$ changes from negative to positive, as observed in the present work.

It is interesting to note that the sign of the magnon polaron anomaly for each mode can change when temperature is varied as discussed above. In addition, due to the independent scattering rates, the sign of TA and LA anomalies can differ at the same temperature. More discussions can be found in Fig. S7(a) [16,37].

Since both the sign and magnitude changes of $\Delta V_{\text{TA-SSE}}$ with varying temperatures primarily depend on the defect scattering of magnons and phonons as well as the four-magnon scattering, the magnon polaron SSE therefore provides a highly sensitive probe to study not only the defect scattering processes of magnons and phonons but also the four-magnon scattering process (based on the exchange interaction) in particular, which normally requires special techniques [38] beyond conventional ones such as thermal conductivity measurements (see SM, Sec. V [16]).

In conclusion, we have investigated the effects of three- and four-magnon scattering processes on the magnon polaron SSE. In YIG crystals with moderately high acoustic quality, the magnon polaron SSE is found to be suppressed at low temperatures but enhanced at high temperatures. The sign change of $\Delta V_{\text{TA-SSE}}$ clearly indicates the critical role of magnon-magnon and in particular four-magnon scattering process in the magnon polaron SSE, and thus provides important insights into the microscopic mechanism of the magnon polaron formation. Understanding and controlling magnon-magnon interaction is of particular interest in other insulating material systems including low-dimensional magnetic structures, antiferromagnets, and other emerging topological magnetic systems in which magnon spectra are significantly different.

Z. S. would like to thank Ding Ding for her assistance on the specific heat and thermal conductivity measurements. Work at Tongji University was supported by National Key R & D Program of China Grand No. 2017YFA0303202, the National Science Foundation of China Grant Nos. 11774259, 12074285, 51671147 and 11874283, 11890703. Work at the University of California Riverside was supported by the US Department of Energy, Office of Science, Basic Energy Sciences under award number SC0012670, and award number DE-FG02-07ER46351.

Z.S. and Q.X. contributed equally to this work.

References:

[1] K. Uchida, S. Takahashi, K. Harii, J. Ieda, W. Koshibae, K. Ando, S. Maekawa, and E. Saitoh, Nature **455**, 778 (2008).
[2] K. Uchida, J. Xiao, H. Adachi, J. Ohe, S. Takahashi, J. Ieda, T. Ota, Y. Kajiwara, H. Umezawa, H. Kawai, G. E. W. Bauer, S. Maekawa, and E. Saitoh, Nat. Mater. **9**, 894 (2010).
[3] C. M. Jaworski, J. Yang, S. Mack, D. D. Awschalom, J. P. Heremans, and R. C. Myers, Nat. Mater. **9**, 898 (2010).
[4] G. E. W. Bauer, E. Saitoh, and B. J. van Wees, Nat. Mater. **11**, 391 (2012).
[5] H. Adachi, K. Uchida, E. Saitoh, and S. Maekawa, Rep. Prog. Phys. **76**, 036501 (2013).
[6] D. Qu, S. Y. Huang, J. Hu, R. Wu, and C. L. Chien, Phys. Rev. Lett. **110**, 067206 (2013).
[7] H. Chang, P. A. P. Janantha, J. Ding, T. Liu, K. Cline, J. N. Gelfand, W. Li, M. C. Marconi, and M. Wu, Sci. Adv. **3**, e1601614 (2017).
[8] T. Kikkawa, K. Shen, B. Flebus, R. A. Duine, K. Uchida, Z. Qiu, G. E. W. Bauer, and E. Saitoh, Phys. Rev. Lett. **117**, 207203 (2016).
[9] B. Flebus, K. Shen, T. Kikkawa, K. Uchida, Z. Qiu, E. Saitoh, R. A. Duine, and G. E. W. Bauer, Phys. Rev. B **95**, 144420 (2017).
[10] L. J. Cornelissen, K. Oyanagi, T. Kikkawa, Z. Qiu, T. Kuschel, G. E. W. Bauer, B. J. van Wees, and E. Saitoh, Phys. Rev. B **96**, 104441 (2017).
[11] R. Ramos, T. Hioki, Y. Hashimoto, T. Kikkawa, P. Frey, A. J. E. Kreil, V. I. Vasyuchka, A. A. Serga, B. Hillebrands, and E. Saitoh, Nat. Commun. **10**, 1 (2019).
[12] R. Schmidt, F. Wilken, T. S. Nunner, and P. W. Brouwer, Phys. Rev. B **98**, 134421 (2018).
[13] T. Liu, W. Wang, and J. Zhang, Phys. Rev. B **99**, 214407 (2019).
[14] S. O. Demokritov, V. E. Demidov, O. Dzyapko, G. A. Melkov, A. A. Serga, B. Hillebrands, and A. N. Slavin, Nature **443**, 430 (2006).
[15] X.-B. Chen, N. T. M. Hien, D. Lee, S.-Y. Jang, T. W. Noh, and I.-S. Yang, New J. Phys. **12**, 073046 (2010).
[16] See Supplementary Material at [url] for more details on the sample fabrication, the characterization of microstructures, the additional data analysis, and which includes Refs. [17-28].
[17] D. Wesenberg, T. Liu, D. Balzar, M. Wu, and B. L. Zink, Nat. Phys. **13**, 987 (2017).
[18] T. B. Noack, H. Y. Musiienko-Shmarova, T. Langner, F. Heussner, V. Lauer, B. Heinz, D. A. Bozhko, V. I. Vasyuchka, A. Pomyalov, V. S. L'vov, B. Hillebrands, and A. A. Serga, J. Phys. Appl. Phys. **51**, 234003 (2018).
[19] K. Mandal, S. Mitra, and P. A. Kumar, EPL Europhys. Lett. **75**, 618 (2006).
[20] W. Kipferl, M. Dumm, P. Kotissek, F. Steinbauer, and G. Bayreuther, J. Appl. Phys. **95**, 7417 (2004).
[21] B. Y. Pan, T. Y. Guan, X. C. Hong, S. Y. Zhou, X. Qiu, H. Zhang, and S. Y. Li, EPL Europhys. Lett. **103**, 37005 (2013).
[22] S. R. Boona and J. P. Heremans, Phys. Rev. B **90**, 064421 (2014).
[23] A. Miura, T. Kikkawa, R. Iguchi, K. Uchida, E. Saitoh, and J. Shiomi, Phys. Rev. Mater. **1**, 014601 (2017).
[24] R. Berman and D. K. C. MacDonald, Proc. R. Soc. Lond. Ser. Math. Phys. Sci. **211**, 122 (1952).

[25] C. T. Walker and R. O. Pohl, Phys. Rev. **131**, 1433 (1963).
[26] M. Sparks, R. Loudon, and C. Kittel, Phys. Rev. **122**, 791 (1961).
[27] P. Pincus, M. Sparks, and R. C. LeCraw, Phys. Rev. **124**, 1015 (1961).
[28] E. Langenberg, E. Ferreiro-Vila, V. Leborán, A. O. Fumega, V. Pardo, and F. Rivadulla, APL Mater. **4**, 104815 (2016).
[29] T. Kikkawa, K. Uchida, S. Daimon, Z. Qiu, Y. Shiomi, and E. Saitoh, Phys. Rev. B **92**, 064413 (2015).
[30] K. B. Modi, M. C. Chhantbar, P. U. Sharma, and H. H. Joshi, J. Mater. Sci. **40**, 1247 (2005).
[31] J. Xiao, G. E. W. Bauer, K. Uchida, E. Saitoh, and S. Maekawa, Phys. Rev. B **81**, 214418 (2010).
[32] H. Adachi, J. Ohe, S. Takahashi, and S. Maekawa, Phys. Rev. B **83**, 094410 (2011).
[33] H. Adachi, K. Uchida, E. Saitoh, J. Ohe, S. Takahashi, and S. Maekawa, Appl. Phys. Lett. **97**, 252506 (2010).
[34] C. M. Jaworski, J. Yang, S. Mack, D. D. Awschalom, R. C. Myers, and J. P. Heremans, Phys. Rev. Lett. **106**, 186601 (2011).
[35] S. M. Rezende, R. L. Rodríguez-Suárez, J. C. Lopez Ortiz, and A. Azevedo, Phys. Rev. B **89**, 134406 (2014).
[36] R. Iguchi, K. Uchida, S. Daimon, and E. Saitoh, Phys. Rev. B **95**, 174401 (2017).
[37] H. Man, Z. Shi, G. Xu, Y. Xu, X. Chen, S. Sullivan, J. Zhou, K. Xia, J. Shi, and P. Dai, Phys. Rev. B **96**, 100406 (2017).
[38] H. Kurebayashi, O. Dzyapko, V. E. Demidov, D. Fang, A. J. Ferguson, and S. O. Demokritov, Nat. Mater. **10**, 660 (2011).